# On the origins of charge transport in spin crossover complexes

Archit Dhingra[1,2]*, and M. Zaid Zaz[3]

[1] Instituto Madrileño de Estudios Avanzados, IMDEA Nanociencia, Calle Faraday 9, 28049 Madrid, Spain

[2] MAX IV Laboratory, Fotongatan 2, Lund, SE-22484, Sweden

[3] Department of Physics and Astronomy, University of Nebraska−Lincoln, Jorgensen Hall, 855 North 16th Street, Lincoln, NE 68588-0299, U. S. A.

**ABSTRACT:** Spin crossover (SCO) complexes are highly promising candidates for a myriad of potential applications in room-temperature electronics; however, as it stands, establishing a clear connection between their spin-state switching and transport properties has been far from trivial. In this Perspective, an effort to unravel the underlying charge transport mechanism in these SCO complexes, via a general theory, is made. The theory presented herein is aimed at providing a unifying picture that explains the widely different trends observed in the spin-crossover-dependent carrier transport properties in the SCO molecular thin film systems.

Spin crossover (SCO) materials, especially the ones with the Fe(II) cation,[1–10] have garnered immense research attention because of the ease with which their spin-states can be reversibly transitioned between the low-spin state (LS) and the high-spin state (HS). Bistability of the spin-states of these systems at room temperature, combined with facile switching between the two (i.e., LS and HS), implies that these materials hold quite a lot of untapped potential for applications in room-temperature electronics. However, to the best of our knowledge, a reliable fundamental understanding of the effects of spin-state switching in the SCO complexes on their carrier transport properties is still absent.[11,12]

Here, a possible qualitative theory elucidating the spin-state-dependent variations in the electrical conductance of the SCO materials is provided. The elements forming the basis of this theory are: (i) activated charge carrier transport; (ii) ligand-to-metal charge transfer (LMCT); (iii) polaron transport. Fathoming the effects of LS ⇌ HS transitions in these SCO systems, with the $[Ar]3d^6$ electronic configuration (Figure 1), on the aforementioned elements is, hence, a viable route towards deciphering their carrier transport mechanism.

In the context of carrier transport in SCO complexes,

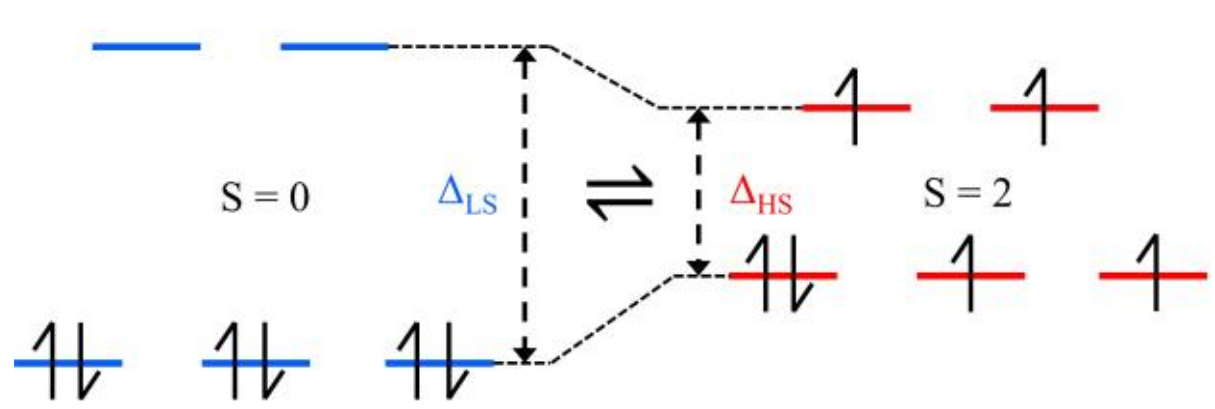

**Figure 1.** $[Ar]3d^6$ electronic configuration of a model SCO complex with Fe(II) cation, in an octahedral ligand field, in the LS and HS states.

it is worth mentioning that attaining the LS ⇌ HS transition by varying the temperature will affect the contribution of activated charge carriers to the transport properties of these materials.[11,13] Position of the Fermi level ($E_F$) with respect to the charge-transfer bands (resulting from the charge transfer between the ligand-like and metal-like molecular orbitals) is bound to alter with the spin-state. As the position of the $E_F$ varies with the spin-state, so does the density of states available for exchange of carriers between the contact and these complexes, translating to a proportionate spin-state dependent change in the conductance of the subject SCO. Reckoning this phenomenon, which has not been articulated in the relevant literature thus far, helps shine light on why charge transfer salts have significant effects on the conductance of these molecules.[14–19]

Moreover, it is now well-established that switching the spin-state of a SCO material leads to a concomitant change in its band gap, which in turn alters its dielectric constant,[1] and the extent of the LMCT within the molecule.[6,7] The strength of the LMCT, which will also depend on the kind of ligand the cation is attached to, and the value of dielectric constant in a given spin-state have direct influence on the formation of polarons in a given SCO system. That is to say, the elongation of bonds upon the spin-state transition from the LS to HS, which weakens the LMCT in these systems, will reduce the strength of orbital overlapping and lower the Debye frequency (phonon frequencies) in these materials; whereas the spin-state dependent variation of the band gap will inversely affect their respective dielectric constant/polarizability.[20] Consequently, such fluctuations will modify the magnitudes of both short-range and long-range electron-lattice interactions, which are known to be responsible for the formation of small and large polarons.[20–24] Small polarons are formed when a self-trapped electronic charge carrier collapses to a single site due to strong short-range electron-lattice interactions, while the formation of large polarons requires long-range Coulombic interactions between an electronic carrier and the lattice.[21,24] Since the mobility of strongly-coupled, coherently moving, large polarons (mathematically represented by the Fröhlich Hamiltonian[25] (see eq 1)) is much higher than that of the incoherently moving small polarons[21] (described by the Holstein Hamiltonian[26,27]), the possibility of their contribution to the transport properties of a given material should not be ignored.

Treating the electrons in first quantization and the phonons in second quantization, the three-term Fröhlich Hamiltonian describing large polarons is given as[25]

$$\hat{H} = \hat{H}_e + \hat{H}_{ph} + \hat{H}_{e-ph}$$
$$= \frac{\hat{p}_e^2}{2m} + \sum_{\boldsymbol{k}} \hbar\omega_{\boldsymbol{k}}( \hat{b}_{\boldsymbol{k}}^{\dagger}\hat{b}_{\boldsymbol{k}} + \frac{1}{2})$$
$$+ \sum_{\boldsymbol{k}} ( V_{\boldsymbol{k}}\hat{b}_{\boldsymbol{k}}^{\dagger}e^{-i\boldsymbol{k}.\hat{\boldsymbol{r}}_e} + V_{\boldsymbol{k}}^{*}\hat{b}_{\boldsymbol{k}}e^{i\boldsymbol{k}.\hat{\boldsymbol{r}}_e}) \quad (1)$$

Here, $\hat{p}_e$ is the momentum operator associated with the electron of mass $m$, $\boldsymbol{k}$ is the phonon momentum, phonon dispersion is denoted by $\omega_{\boldsymbol{k}}$, and $V_{\boldsymbol{k}}$ is the electron-phonon interaction strength. Specific forms of these functions are dictated by the system of interest.[22,28] $\hat{b}_{\boldsymbol{k}}^{\dagger}$ and $\hat{b}_{\boldsymbol{k}}$ are the creation and annihilation operators corresponding to a phonon with wave number $\boldsymbol{k}$.

Now, all things considered, the dilemma pertaining to the different trends observed in spin-state-dependent carrier transport can be explained by examining the exact interplay of all the above-mentioned elements for the SCO complexes of interest. To elaborate further, studies indicating higher electrical conductivity for particular SCO complexes in the LS[11,13,29] can now be reconciled with the studies suggesting higher electrical conductance for a different set of SCO materials in the HS[2,30] by understanding the contribution of each of the aforementioned elements of our theory. Observation of higher electrical conductivity in the LS[11,13,29] can be explained as follows: the lattice parameters in the LS are shorter than in the HS, implying stronger LMCT due to stronger orbital overlap. As a result, the Debye temperature[31] of the SCO material in the LS would be higher than its Debye temperature in the HS, reducing the number of phonons in the LS when compared to the HS. Therefore, in this case, the main contributions to the carrier transport would come from the thermally activated charge carriers and large polarons as the small-polaron transport requires assistance from phonons.[21] On the other hand, higher electrical conductance in the HS[2,30] of a given SCO complex can be understood as a combined result of small-polaron hopping, because the elongation of the bonds[7] will favor strong short-range electron-lattice interactions over long-range ones, in addition to thermally activated charge carriers. This is particularly consistent with the considerable strength of onsite Hubbard correlation in the LS of one such complex,[32] which depicts higher electrical conductance in its HS.[2,30] Besides, since the effective masses of both kinds of polarons are huge compared to that of thermally activated charge carriers, large currents would indicate that the given material possesses quite a lot of charge carriers that can be thermally activated.[21]

To summarize, this work aims to provide a qualitative theory to gather physical insights into the underlying charge transport mechanism in the SCO complexes. The quantitative understanding of the carrier transport in these systems would rely on the density of activated charge carriers that can be thermally activated, their respective activation barriers, the kind of ligand the cation is attached to, the type of polaronic transport, and the concentration of phonons; and will, therefore, be material specific. Nevertheless, our general theory helps with the explanation and reconciliation of conflicting trends observed in the spin-crossover-dependent carrier transport properties in the SCO materials.[2,11,13,29,30]

## AUTHOR INFORMATION

### Corresponding Author

*Archit Dhingra, e-mail: archit.dhingra@maxiv.lu.se

### Conflicts of interest

There are no conflicts of interest to declare.

## ACKNOWLEDGMENTS

The authors thank Prof. Peter A. Dowben (Department of Physics & Astronomy, UNL) for fruitful discussions. A.D. acknowledges the financial support from the Spanish Ministry of Science and Innovation through "Severo Ochoa" (Grant CEX2020-001039- S) and "María de Maeztu" (Grant CEX2018-000805-M) Programmes for Centres of Excellence in R&D. Financial support by the Comunidad de Madrid through (MAD2D-CM)-MRR MATERIALES AVANZADOS-IMDEA-NC is also acknowledged.